\documentclass{iau}

\usepackage{amsmath}
\usepackage{graphicx}
\usepackage{multirow,multicol}
\usepackage{newtxmath}

\begin{document}

\lefttitle{Crowther, P.~A. \& Bestenlehner, J.~M.}
\righttitle{New Pipeline for Efficient Analysis of Large Spectroscopic Samples of OB Stars}

\jnlPage{1}{7}
\jnlDoiYr{2025}
\doival{10.1017/xxxxx}

\aopheadtitle{Proceedings IAU Symposium}
\volno{402}
\editors{A. Wofford,  N. St-Louis, M. Garcia \&  S. Simón-Díaz, eds.}

\title{New Pipeline for Efficient Analysis of Large Spectroscopic Samples of OB Stars}

\author{Paul~A.~Crowther$^{1}$, Joachim~M.~Bestenlehner$^{1,2}$}
\affiliation{1: Astrophysics Research Cluster, School of Mathematical \& Physical Sciences, University of Sheffield, Hicks Building, Hounsfield Road, Sheffield S3 7RH, UK\\
2: School of Chemical, Materials and Biological Engineering, University of Sheffield, \\ Sir Robert Hadfield Building, Mappin Street, Sheffield S1 3JD, UK}

\begin{abstract}
We present a new spectroscopic pipeline designed to analyse large numbers of hot massive stars homogeneously. The pipeline has been developed to utilise large grids of FASTWIND non-LTE, line blanketed models in which spherical geometry is 
adopted, and uniquely incorporates model errors. The pipeline has been applied to three contemporary datasets involving Very Large Telescope spectroscopy of OB stars in the Magellanic Clouds, namely the VLT FLAMES Tarantula Survey (VFTS), XShooting-ULLYSES (XShootU) and Binaries at Low Metallicity (BLOeM). We find satisfactory agreement with previous detailed temperatures and surface gravities, although strong nebular contamination, binarity and  disk emission from OBe stars are problematic for automatic pipelines,
requiring visual inspection of fits. The tool has been incorporated into the pipeline for the VISTA/4MOST pipeline.
\end{abstract}


\maketitle

\section{Introduction}

Efficient analysis of large samples of spectroscopic datasets is widespread for late-type stars \citep[e.g. The Payne,][]{ThePayne}, owing to the 
availability of computationally inexpensive plane parallel, line blanketed, LTE models \citep[e.g. ATLAS12,][]{ATLAS12}.
The situation is a little more challenging for non-supergiant B stars requiring non-LTE models 
\citep[e.g. TLUSTY,][]{LanzHubeny2007}, although hybrid approaches involving LTE atmospheric structure models 
supplemented by non-LTE line formation are also in common use \citep[see e.g.][]{NievaPrzybilla2007}. However, the situation for 
O stars is problematic, owing to the requirement of spherical geometry, line blanketing and non-LTE. 

Several model atmosphere codes have been designed for hot luminous stars with strong outflows, although significant resources are required for 
individual CMFGEN \citep{Hillier-Miller1998} or PoWR \citep{Sander+2015} calculations, although these are reduced somewhat
with FASTWIND \citep{Puls+2005}. For this reason, moderate  samples of OB stars are usually tackled using $\chi^{2}$ minimalization approaches exploiting pre-existing FASTWIND grids
\citep[e.g. IACOB-GBAT,][]{IACOB-GBAT} or a Genetic Algorithm using natural selection \citep[e.g. KIWI-GA,][]{Brands+2022}. 

Current (e.g. VLT/FLAMES) and upcoming 
(e.g. VISTA/4MOST) highly mutiplexed spectrographs necessitate a different approach. 
Spectroscopic pipelines for cool stars have been extended to OBA stars \citep[e.g. HotPayne,][]{HotPayne}, albeit retaining plane-parallel LTE ATLAS12 models, which
are ill suited to blue supergiants or O stars.  Abdul-Masih (these proceedings) 
discusses the outlook for the use of
Neural Networks to  emulate FASTWIND spectroscopic analyses. Here, we consider an alternative approach, involving a new spectroscopic pipeline coupled
to a large grid of FASTWIND models to analyse large samples of OB stars, and provide some pilot studies of Magellanic Cloud hot massive stars.

\section{Pipeline}

A full description of the pipeline is provided by \citet{Bestenlehner+2024}. In brief, observations across the complete spectral range, including error spectra, are compared to synthetic spectra from
grids of FASTWIND models using a $\chi^{2}$ minimisation
\begin{equation*}
    \chi^2 = (\vec{d} - \mathrm{R}\vec{s})^{\mathrm T}\mathrm{N}^{-1}(\vec{d} - \mathrm{R}\vec{s})
\end{equation*}
with $\vec{d}$ the observed and $\vec{s}$ the synthetic spectra, $\mathrm{R}$ the instrumental response matrix and observational, diagonal error matrix $\mathrm{N}$. Model uncertainties are budgeted into the parameter determination. 
The goodness of fit is usually evaluated by calculating the reduced $\chi^{2}$, which uses in our case the diagonal of the error covariance matrix.

FASTWIND (v10.6) models include H, He, C, N, O, Si and Mg as explicit elements at the relevant (LMC or SMC) metallicity. Grids cover the following parameter space $\log T_{\rm eff}$ (K) over [4.0, 4.775] in 0.025 dex steps, 
$\log g$ (cm\,s$^{-2}$) over [1.5, 4.5] in 0.2 dex steps, wind-strength parameter $\log Q$ ranging from --11.4 to --15.0 in 0.3 dex steps\footnote{$Q = \dot{M} (R_{\ast} v_{\infty})^{-3/2}$ with units $M_{\odot}$\,yr$^{-1}$, $R_{\odot}$ and 
km\,s$^{-1}$.} and Helium abundances in mass-fraction $Y$ over [0.15, 0.55] in 0.05 steps. A smooth wind with volume filling factor $f_{\rm v} = 1 $ and $\beta = 1$ velocity law was assumed and the micro-turbulent velocity 
was uniformly set to $\varv_{\rm mic}=10$ km\,s$^{-1}$ in the model grids. We convolved our synthetic grid with a fixed $\varv_{\rm mac} = 20$ km\,s$^{-1}$ and assumed any additional broadening is due to rotation, with 
projected rotational velocities of $\varv_{\rm e} \sin i = [0, 10, 20, 35, 50, 75, 100, 150, 200, 250, 300, 350, 400, 450, 500]$ km\,s$^{-1}$.

\begin{figure}[htbp]
\centering
\includegraphics[width=0.4\columnwidth]{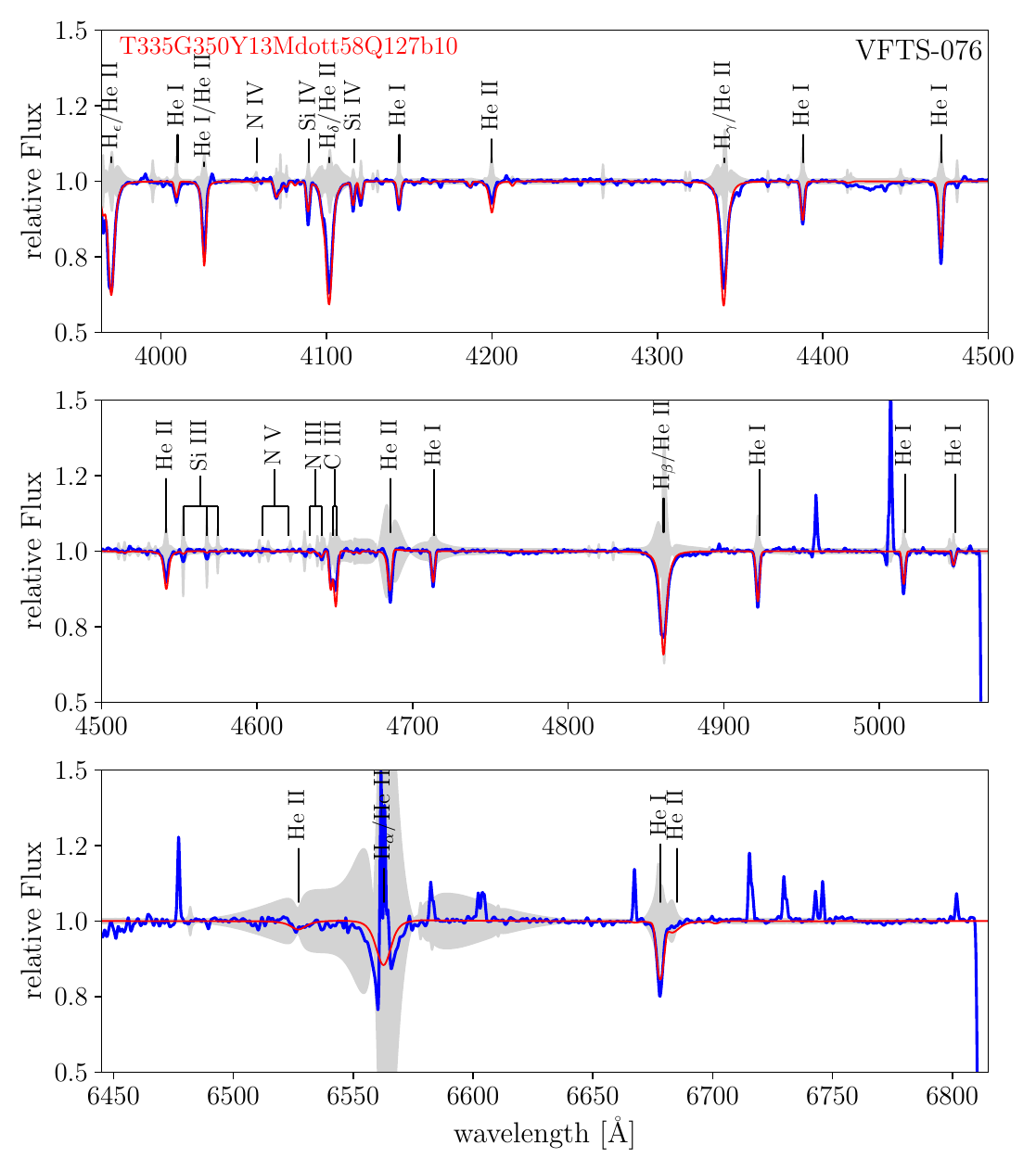}
\includegraphics[width=0.59\columnwidth]{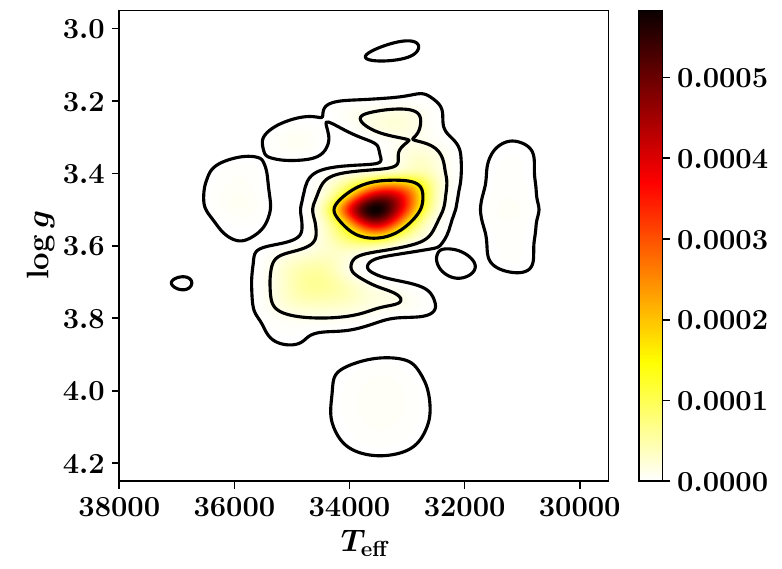}
  \caption{Left: Spectroscopic fit (red) to observations (blue) for VFTS 76 \citep[O9.2\,III,][]{Evans+2011} from \citet{Bestenlehner+2024}. The grey shaded region is the square root of the diagonal elements of the error uncertainty matrix. Right: Probabilities heatmap of pipeline solution to VFTS 076 in the $\log g - T_{\rm eff}$ plane. Contours indicate 2D confidence intervals of 39.4, 86.5 and 98.9 per cent.}
  \label{VFTS076}
\end{figure}

Once setup, the pipeline has been designed to rapidly analyse large numbers  (hundreds to thousands) of spectroscopic datasets automatically, in contrast to interactive star-by-star processing with IACOB-GBAT or KIWI-GA. The runtime scales exponentially with the number of spectra, since all stars are analysed simultaneously so that the model error can be determined. Nevertheless, 1000 star can be analysed in less than 2 hours. Consequently, for very large
datasets, it would be prudent to sort these into broadly similar objects, which likely possess comparable model errors given their parameter space. However, once a converted error uncertainty matrix is obtained, the matrix operations can be limited to the $\chi^{2}$-minimization, and switched to a star-by-star analysis. In the next section we apply the pipeline to several medium to large OB datasets, some of which have previously been analysed using conventional techniques.

\section{Application to large spectroscopic datasets}

In order to assess the validity of the spectroscopic pipeline, we have applied it to  contemporary datasets involving OB stars in the Magellanic Clouds.

\subsection{VFTS}

The VLT-FLAMES Tarantula Survey \citep{Evans+2011} used the FLAMES instrument at the Very Large Telescope (VLT) to obtain multiple-epoch, intermediate resolution (blue and H$\alpha$) spectroscopy of 
nearly 1000 OB stars within the Tarantula Nebula of the LMC. We have applied the pipeline to 240 single O stars \citep{Bestenlehner+2024}, the majority of which have previously been analysed by \citet{Sabin-Sanjulian+2017} and \citet{Ramirez-Agudelo+2017}. 

Overall, reasonable agreement between inferred temperatures and surface gravities were obtained, although strong nebular contamination is problematic for a sizeable subset of the VFTS O star sample, as illustrated in the left panel of Fig.~\ref{VFTS076} for
VFTS 76 (O9.2\,III) from  \citep{Bestenlehner+2024}. The right panel of Fig.~\ref{VFTS076}  shows a  probability heatmap of  temperature versus gravity for VFTS 76 showing a relatively clean locus, albeit with the possibility of a higher gravity owing to the degeneracy between surface gravity and mass-loss, since increased mass-loss fills the wings of Balmer lines.

\begin{figure}[htbp]
\centering
\includegraphics[width=0.505\columnwidth]{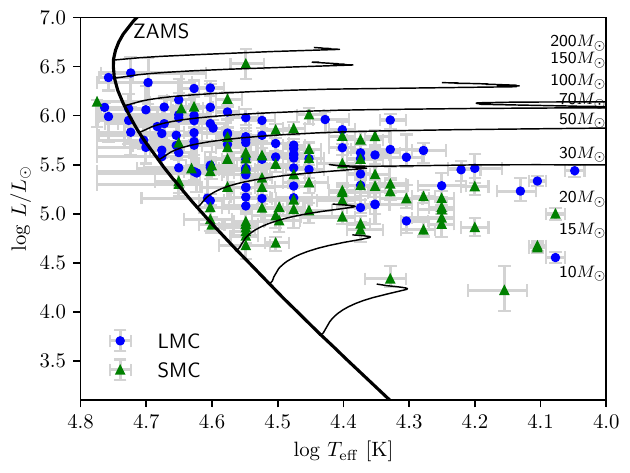}
\includegraphics[width=0.485\columnwidth]{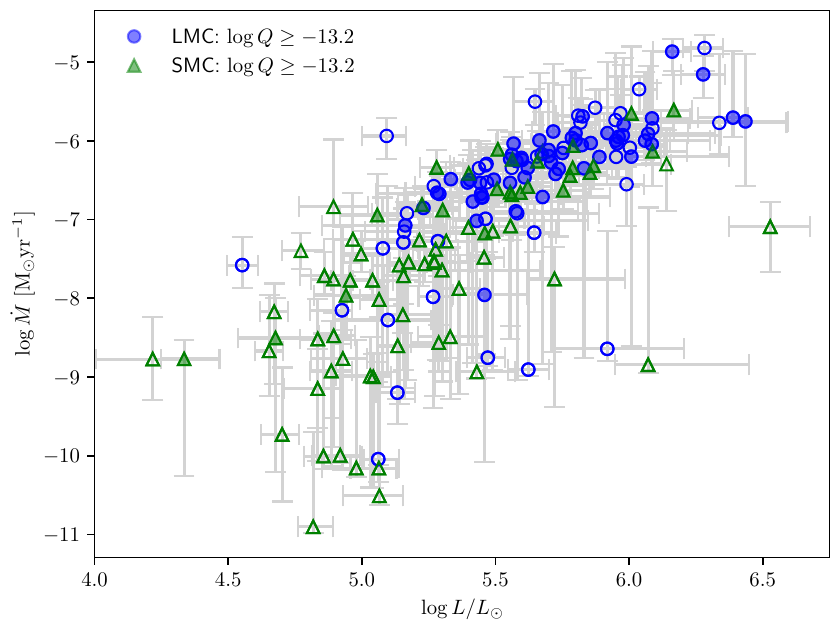}
  \caption{Left: Pipeline-derived Hertzsprung-Russell diagram of XShootU OB stars in the LMC (blue circles) and SMC (green triangles), together with LMC evolutionary tracks for main sequence stars from \citet{Brott+2011} and \citet{Kohler+2015}; Right: 
  Pipeline mass-loss rates of LMC (blue circles) and SMC (green triangles) XShootU OB stars versus luminosity from \citet{Bestenlehner+2025a}, in which stars with weak wlnds are indicated with open symbols. }
  \label{XShootU}
\end{figure}

\subsection{XShootU}

The STScI Director's Discretionary programme ULLYSES \citep{Roman-Duval+2025}  includes medium resolution far-UV spectroscopy for over 200 Magellanic Cloud OB and Wolf-Rayet stars, and has been complemented by intermediate resolution VLT/Xshooter optical spectroscopy, via XShooting-ULLYSES
\citep[XShootU,][]{Vink+2023}. We have analysed a subset of 97 LMC and 77 SMC OB stars from XShootU, excluding unambiguous SB2 binaries, OBe stars and Wolf-Rayet stars \citep{Bestenlehner+2025a}. Temperatures are broadly in line with previous 
spectroscopic studies, although early O stars proved problematic owing to poor normalization of spectra in the vicinity of the N\,{\sc v} $\lambda\lambda$4603-20 doublet and He\,{\sc ii} $\lambda$4686 which resulted in a lower weight for this wavelength range when minimising the $\chi^{2}$. 

The pipeline-derived Hertzsprung-Russell diagram for XShootU OB stars is presented in the left panel of Fig.~\ref{XShootU}, highlighting a deficiency of hot, luminous OB stars in the SMC. Evolutionary masses were obtained from the Bayesian inference method BONNSAI \citep{Schneider+2014} coupled with evolutionary models from \citet{Brott+2011} and \citet{Kohler+2015} for main sequence stars, or \citet{Schootemeijer+2019} for post-main sequence stars based on Terminal Age Main Sequence luminosities. Median O-type (B-type) masses are 32.3 $M_{\odot}$ (19.6 $M_{\odot}$) in the SMC, versus 46.0 $M_{\odot}$ (27.3 $M_{\odot}$) in the LMC. 

UV spectroscopy from ULLYSES is required to determine terminal wind velocities \citep[e.g.][]{Hawcroft+2024} and mass-loss rates for stars with weak winds, but the inclusion of H$\alpha$ in XShootU datasets permits comparisons between LMC and SMC OB stars with dense winds ($\log Q \geq -13.2$). OB stars in the LMC sample exhibit somewhat higher mass-loss rates than their SMC counterparts, as illustrated in the right panel of Fig.~\ref{XShootU}.

\begin{figure}[htbp]
\centering
\includegraphics[width=0.49\columnwidth]{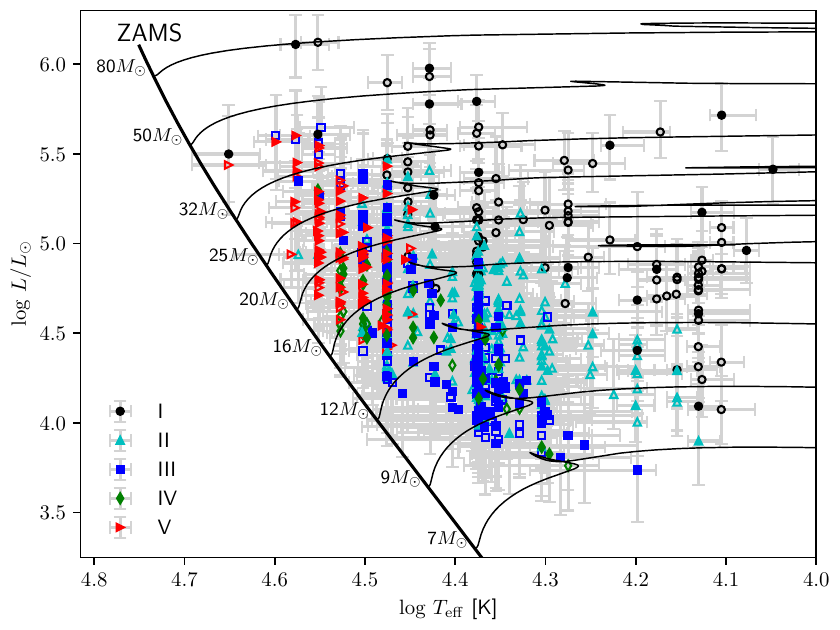}
\includegraphics[width=0.49\columnwidth]{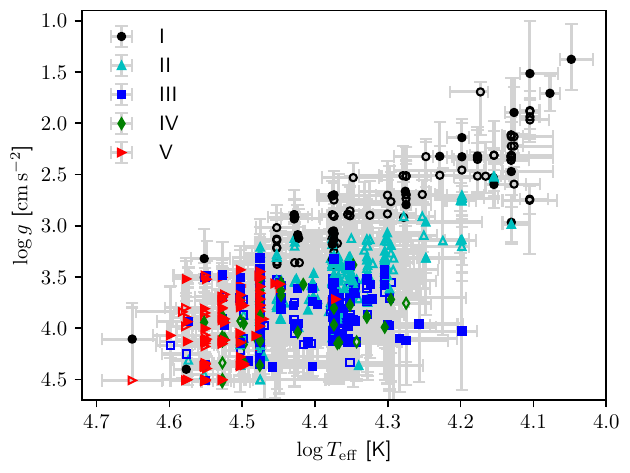}
  \caption{Left: Pipeline-derived Hertzsprung-Russell diagram of BLOeM OB stars in the SMC, colour coded by luminosity class, with SMC evolutionary tracks from \citet{Schootemeijer+2019}. Right: Comparison between effective temperatures and surface gravities of BLOeM OB stars (Kiel diagram). Open symbols are single according to analysis of the initial 9 epoch dataset, filled symbols are multiple. }
  \label{BLOeM}
\end{figure}

\subsection{BLOeM}

The Binaries at Low Metallicity (BLOeM) survey  involved 25 epochs of intermediate resolution blue optical spectroscopy of 929 massive stars in the SMC using the FLAMES instrument at the VLT \citep{Shenar+2024}. We have analysed a subset of 778 OB stars, using the cross-correlated and co-added first 9 epoch datasets, excluding unambiguous SB2 systems, OBe stars and AF supergiants \citep{Bestenlehner+2025b}.  The limited spectral range of $\lambda\lambda$3964-4567 sampled by BLOeM provided most key spectroscopic diagnostics, aside from wind diagnostics (H$\alpha$, He\,{\sc ii} $\lambda$4686)\footnote{Following the completion of the study, single-epoch He\,{\sc ii} $\lambda$4686 and H$\alpha$ spectroscopy was obtained for all BLOeM targets (Mahy, priv. comm.)}. An increased weighting was assigned to Si\,{\sc iv} $\lambda$4089 in order to avoid an unphysical gap in solutions for early B stars close to $\sim$25 kK. 

Although efforts were made to exclude OB stars with  strong nebular or disk emission from the sample, those with weaker (Balmer) line emission will have been retained, affecting surface gravity and wind density determinations. Temperatures are broadly in agreement with literature results for stars in common. A subset of BLOeM O and early B stars have been analysed with IACOB-GBAT, revealing slightly lower  pipeline temperatures and surface gravities, though higher Helium abundances and rotational velocities.
The pipeline-derived Hertzsprung-Russell diagram (Kiel diagram) of BLOeM OB stars is presented in the left (right) panel of Fig.~\ref{BLOeM}. Open symbols are single according to analysis of the first 9 BLOeM epochs, with filled symbols multiple. It is apparent that OB stars from BLOeM represent a very different population from those included in ULLYSES/XShootU from comparison with Fig.~\ref{XShootU}. We recover the anticipated lower mass cut-off at 8 $M_{\odot}$ from the survey design, using a Bayesian inference method coupled to  SMC metallicity evolutionary models \citep{Brott+2011}, with median masses of 19.8 $M_{\odot}$ (12.6 $M_{\odot}$) for O-type (B-type) stars.

\section{Conclusions}

Robust spectroscopic analysis of very large samples of O and early B stars remains challenging since current approaches such as IACOB-GBAT \citep{IACOB-GBAT} or KIWI-GA \citep{Brands+2022} require sizeable (human and computational) resources.
We have developed a spectroscopic pipeline which has been designed to efficiently, and homogeneously, obtain physical properties for large numbers of OB stars, taking into account model errors. 
Application to three moderate spectroscopic samples (VFTS, XShootU, BLOeM) reveals
overall satisfactory temperatures and surface gravities with respect to previous studies, although binarity and/or strong nebular/disk emission remain problematic for automated treatment, so visual inspection of fits is still required. Our tool has been incorporated into the 
4MOST pipeline (providing $T_{\rm eff}, \log g$, wind density) although individual star-by-star analysis is still required for UV spectral fitting and/or detailed metal abundances in view of the parameter space involved.

\section*{Acknowledgements}

PC acknowledges financial support from the School of Mathematical and Physical Sciences at the University of Sheffield which permitted in-person attendance at the Symposium.

\bibliographystyle{iaulike}

\end{document}